\documentclass[%
 aip,
 pop,
 amsmath,amssymb,
 reprint,%
english, 
]{revtex4-1}

\usepackage{graphicx}
\usepackage{dcolumn}
\usepackage{bm}

\usepackage{float} 
\usepackage{babel} 
\usepackage{textgreek} 
\usepackage[margin=0.79in]{geometry}
\newcommand{\um}{\textmu m} 

\begin{document}

\preprint{AIP/123-QED}

\title{Backward-propagating MeV electrons from $10^{18}$ W/cm$^2$ laser interactions with water}

\author{J.T. Morrison}%
 \affiliation{Fellow, National Research Council , Washington, D.C., 20001, USA}
 \author{ E.A. Chowdhury}%
 \affiliation{\mbox{Department of Physics, The Ohio State University, Columbus, OH, 43210, USA}}
  \affiliation{Intense Energy Solutions, LLC., Plain City, OH, 43064, USA}
 
 \author{K.D. Frische}%
  \affiliation{Innovative Scientific Solutions, Inc., Dayton OH, 45459, USA}
 
\author{S. Feister}
 \email{feister.7@osu.edu}
 \affiliation{\mbox{Department of Physics, The Ohio State University, Columbus, OH, 43210, USA}}
 \affiliation{Innovative Scientific Solutions, Inc., Dayton OH, 45459, USA}

\author{V.M. Ovchinnikov}
 \affiliation{Innovative Scientific Solutions, Inc., Dayton OH, 45459, USA}
 
\author{J.A. Nees}%
 \affiliation{\mbox{Center for Ultra-fast Optical Science, University of Michigan, Ann Arbor, MI, 48109, USA}} 
 \affiliation{Innovative Scientific Solutions, Inc., Dayton OH, 45459, USA}

\author{C. Orban}%
 \affiliation{\mbox{Department of Physics, The Ohio State University, Columbus, OH, 43210, USA}}
 \affiliation{Innovative Scientific Solutions, Inc., Dayton OH, 45459, USA}

\author{R.R. Freeman}
 \affiliation{\mbox{Department of Physics, The Ohio State University, Columbus, OH, 43210, USA}}
 
\author{W.M. Roquemore}
 \affiliation{Air Force Research Laboratory, Dayton, OH, 45433, USA}


\date{\today}

\begin{abstract}
We present an experimental study of the generation of $\sim$MeV electrons opposite to the direction of  laser propagation following the relativistic interaction at normal incidence of a $\sim$3~mJ, \mbox{$10^{18}$ W/cm$^2$} short pulse laser with a flowing 30 \um{} diameter water column target. Faraday cup measurements record hundreds of pC charge accelerated to energies exceeding 120~keV, and energy-resolved measurements of secondary x-ray emissions reveal an x-ray spectrum peaking above 800~keV, which is significantly higher energy than previous studies with similar experimental conditions and more than five times the $\sim$110~keV ponderomotive energy scale for the laser. We show that the energetic x-rays generated in the experiment result from backward-going, high-energy electrons interacting with the focusing optic and vacuum chamber walls with only a small component of x-ray emission emerging from the target itself. We also demonstrate that the high energy radiation can be suppressed through the attenuation of the nanosecond-scale pre-pulse. These results are supported by 2D Particle-in-Cell (PIC) simulations of the laser-plasma interaction that exhibit beam-like backward-propagating MeV electrons.
%
\end{abstract}

\pacs{41.75.Jv, 41.75.Ht, 41.50}
\maketitle
%

\section{\label{sec:Introduction}Introduction}

Laser plasma interactions (LPI) in the relativistic regime have been a topic of major experimental and theoretical studies over the last two decades. These efforts are motivated by the desire to investigate the fundamental science involved in these laser interactions \cite{koenig2005, akli2008, wang2013} and by the possibility of developing a number of applications. These applications include novel particle sources such as relativistic electron \cite{leemans2006, esaray2009} and ion \cite{yin2006} beams and monochromatic x-ray and gamma-ray generation \cite{mackenroth2011}. Laser-accelerated electrons may also find use in radiotherapy \cite{glinec2006, fuchs2009} and other applications may emerge by creating highly-pulsed x-ray and ultraviolet radiation through Compton scattering. Many groups are currently pursuing laser-accelerated electron sources with these goals in mind \cite{zulick2013, rao2007, fullagar2007, hatanaka2008, mordovanakis2009}.

Previous experiments have shown that electrons can be accelerated to hundreds-of-keV energies with lasers of only a few mJ energy when focused to moderately-relativistic intensities. \citet{uhlig2011} used a double-pulsed beam to demonstrate that laser intensities in the range \mbox{10$^{15}$ - }\mbox{10$^{18}$ W/cm$^2$} impinging on a flowing water jet can produce forward-going electron beams with kinetic energies between 280 and 390~keV.  Observing the most-enhanced acceleration at oblique incidence, \citeauthor{uhlig2011} focused their study on obliquely incident LPI and the subsequent forward-directed secondary x-rays produced from accelerated electrons. They did note that in the case of normal incidence, significant x-ray emissions in the forward direction could be obtained, but only when using a double pulse. They suggested that this effect of normal incidence might be of interest for future investigation.

\begin{figure*}
\centering
 \includegraphics[width=17cm]{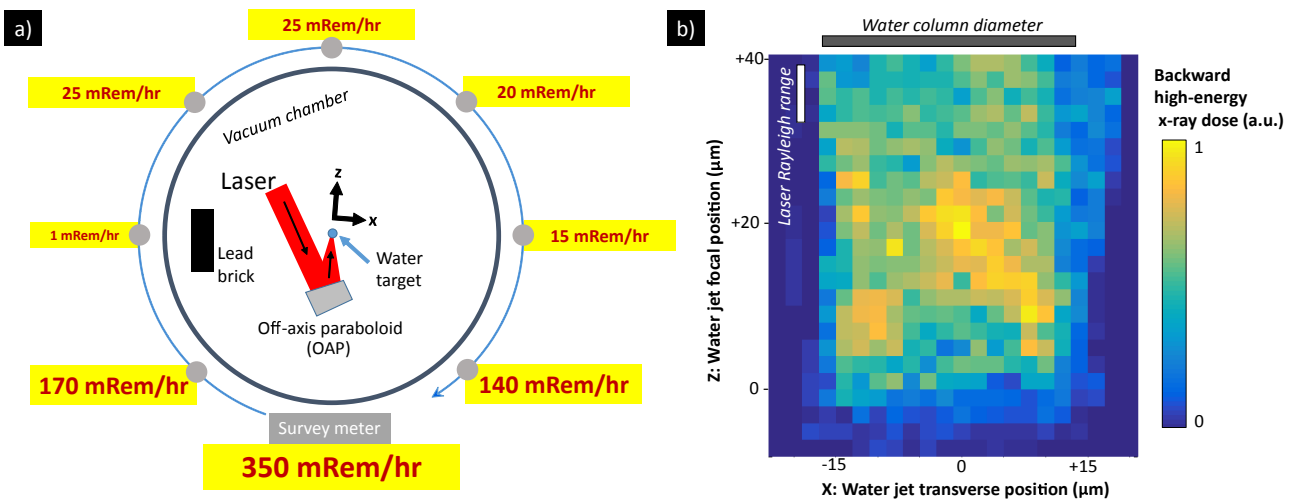}
 \vspace{-0.3cm}
 \caption{\label{fig:dose} a) Survey meter measurements taken 60 cm from the target (30 cm outside the vacuum chamber), on the laser axis and also every 45 degrees, are shown for normal-incidence laser interactions with polarization along the indicated X direction. The measurements highlighted in panel (a) show that at these conditions most of the radiation is directed opposite to the laser propagation.
 b) A plot of the strength of backward-going radiation as one translates the water column (\emph{not} the laser focus) in the Z and X directions indicated in panel (a). X represents water jet transverse position where 0 indicates a normal incidence and $\pm$15~\um{} indicates focusing on the curved edge of the 30~\um{} diameter water column. Z represents water jet focal depth position, where values increase as the water jet moves away from the OAP and 0 represents coincidence of best laser focus with water jet surface at X$=$0. X$=$0 is accurate to $\sim$3~\um{} and Z$=$0 is accurate to $\sim$10~\um{} (see text). The color axis shows the high-energy x-ray dose reading from an under-biased Amptek GAMMA-RAD5 detector placed 30 cm behind the vacuum chamber (in the position of the gray rectangle of panel (a)). Each pixel represents the average of 67 shots. Note that the emission is substantial only when the water surface is beyond the focus point.} 
\end{figure*}

Here we describe experiments with a water target at normal and oblique incidence in which an unexpectedly large radiation dose is recorded at many meters from the interaction, and peaked not in the forward direction, but in the backward direction. Further, we present and describe large scale PIC simulations which support these observations. Dosimetry measurements (see Fig. \ref{fig:dose}a) show that the highest radiation dose is in the backward direction, \textit{opposite} the incident laser propagation. Fig. \ref{fig:dose}b shows that secondary backward-going radiation is high for laser interaction along the edges of the water column when the water column front center surface is about 10~\um{} beyond laser focus. However, the highest dose comes from normal incidence, when the water column is placed about 20~\um{} beyond the focus. Little or no dose is measured when the front surface of the water column is moved to a position nearer to the off-axis parabolic mirror (OAP), before laser focus.

This phenomenon is not subtle: the dose is more than ten times greater in the backward direction than in the forward direction, and is surprisingly intense. We have imaged backward-going electrons onto a Lanex phosphor screen and shown that they shift in the presence of magnetic fields, in a manner consistent with beam-like electron radiation; both of these measurements of high energy, backward-flowing, beam-like electrons are discussed in detail by \citet{orban2015}

In this work, we concentrate on another remarkable phenomenon associated with the recorded dose: the existence of high energy x-rays with peak energies on the order of 1~MeV propagating counter to the laser along target normal and arising from the interaction of less than 3~mJ of short-pulse laser light focused to \mbox{$10^{18}$ W/cm$^2$} on a water target. These x-rays measured far from the target have two sources: the first is within the target itself, and the second is through electron collisions with the gold-coated aluminum OAP focusing optic and steel vacuum chamber walls. We give a general experimental description of the primary backward-going electron radiation, but we did not measure the electron energy spectra because the experiment is performed at normal incidence to the target and the electrons are radiated backward; while possible with future significant modifications, a typical electron spectrometer implementation would block the incoming laser. Here, we analyze the energy spectrum of secondary x-rays and find that the x-ray spectrum emitted isotropically from the target differs significantly from the spectrum emitted in the backward direction, having a much lower intensity. We find that the generation of radiation in the backward direction is crucially dependent on the level of laser pre-pulse, an experimental result that is corroborated by PIC simulations. The importance of finite laser pre-pulse in this experiment complements efforts in pursuit of new regimes in ion acceleration, where the use of thinner and structured targets often requires control or inhibition of laser pre-pulse \cite{henig2009enhanced,jiang2014effects, antici2007,fuchs2006,jung2011}.

\section{Experimental Setup}\label{sec:Setup}

\begin{figure*}
\centering
 \includegraphics[width=17cm]{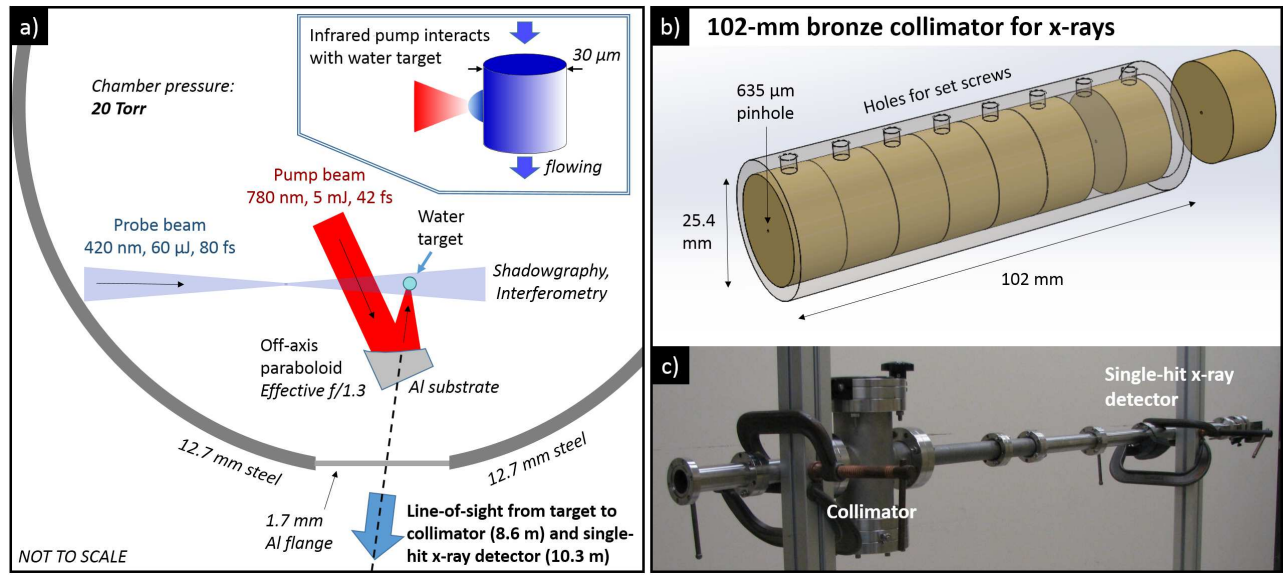}
 \caption{\label{fig:setup} a) A more detailed schematic of the experimental setup than shown in Fig. \ref{fig:dose}a. As in Fig. \ref{fig:dose}a, the red pump beam irradiates the water jet at normal incidence with polarization perpendicular to the water column, producing backward-going electrons that scatter with the focusing element to produce secondary x-rays. A variable-delay, short-pulse probe beam images the interaction region to provide shadowgraphy and interferometry of the pre-pulse interaction with the target and the longer-timescale hydrodynamic reaction to the pump pulse \cite{feister2014}. An Amptek X-123 single-hit x-ray spectrometer \cite{Amptek123_06} is located 10~m from the target. The device is aligned to collect x-rays originating from the back of the OAP and the target itself. b) The signal-to-noise of the single-hit x-ray detector is enhanced with a custom-designed bronze collimator. This collimator consists of a stack of eight bronze discs each containing a 0.025-in. pinhole to allow only a narrow acceptance angle for x-rays. c) The single-hit x-ray detector and its collimator are mounted 1.7~m apart on opposite sides of a vacuum tube to further improve the signal-to-noise ratio for detecting line-of-sight x-rays. Monte Carlo (MCNP) modeling of line-of-sight x-rays and contamination from scattered secondaries up to 2.5~MeV predicts a signal-to-noise ratio better than 10:1. }
\end{figure*}

The experimental study described here utilized the modified Red Dragon laser (KM Labs) at the Air Force Research Laboratory (AFRL) at Wright-Patterson Air Force Base. Fig. \ref{fig:setup}a shows the experimental setup in detail with an inset schematic of the laser-target interaction. The Red Dragon is a Ti: sapphire based 1 kHz system producing \mbox{1 - 5 mJ}, 42~fs Gaussian full width at half maximum (FWHM) pulses at 780~nm~$\pm$~20~nm. The final focusing optic is an aluminum-substrate, gold-coated $30 ^{\circ}$ OAP (Edmund Optics 63-192, effective $f/1.3$) producing a measured 2.2~\um{} FWHM focal spot and a peak laser intensity of \mbox{$1 \times 10^{18}$ W/cm$^2$} in vacuum.

The laser intensity was determined through a series of measurements at full amplification, with reflective ND filters in place to attenuate the beam. A 42 fs pulse duration (FWHM) was measured using a single-shot second-order autocorrelator \cite{salin1987} (Coherent), assuming a Gaussian temporal profile. A 2.2~\um{} FWHM focal spot was measured in-situ using a 20x Edmund Optics 59-878 microscope objective and a 2D Gaussian fit to the image. To calculate the energy in the laser focus, an energy meter (Coherent LabMAX-TOP / PM150-50C) was used to measure pulse energy prior the final focusing OAP, and then this number was reduced according to two sources of energy loss. First, about 45\% of the energy is lost due to angular scattering from imperfections in the machine-grooved, gold-coated OAP. This loss was determined by using the laser focus to bore a $\sim$300~\um{} hole in aluminum foil (a hole area much larger than the actual laser focus), then reducing the energy to avoid air breakdown and comparing energy meter values before the OAP and after the laser focus. Second, only 70\% of the energy reaching the imaging apparatus appears in the Gaussian focus. This second ratio was determined by assessing the image pixel-value integration of the 2D Gaussian fit and comparing it to the integration of the entire background-reduced focal spot image. From these calculations we have found that when 5~mJ is incident on the OAP, only about 2.8~mJ reaches the focal area and 2~mJ is contained in the 2.2~\um{} Gaussian focus. From these numbers we obtain a peak focal intensity of \mbox{$1 \times 10^{18}$ W/cm$^2$}.

The target was a 30~\um{} diameter vertical laminar distilled water column flowing from a 30~\um{} glass capillary nozzle. The water column flows at 14~m/s and remains laminar for several mm after exiting the nozzle, and its position remains stable to within 2~\um{}. A background vapor pressure of 20~Torr was maintained to prevent freezing of the flow.  The laser is polarized perpendicular to the water flow direction, and comes to a focus 10 to 20~\um{} in front of the liquid water surface as was seen by examining air breakdown at 200~Torr in shadowgraphy. Fig. \ref{fig:dose}b shows a plot of high-energy x-ray dose as water column transverse and focal position are scanned; all other data presented in this paper were taken at normal incidence. To ensure normal incidence, the laser was vertically squared with the water column (verified in shadowgraphy) and horizontally centered on the water column curvature (verified using a microscope objective to within 3~\um{}). Probe beam shadowgraphy indicates that material disturbed by interaction with the 1 kHz laser fully exits the interaction region within $\sim$30~\textmu{}s \cite{feister2014}, and so the experiment can be run continuously at 1 kHz.

Ultra-intense laser experiments with near-solid density targets at normal incidence present a challenge: optics in the laser can be damaged by the reflected light from the plasma critical density surface that can be amplified as it travels backward through the laser amplifier chain. To avoid damage due to normal incidence reflections, the laser system was modified by optically isolating the amplifiers using polarizers and Pockels cells. A 20-mm aperture Pockels cell (FastPulse Lasermetrics 5406SC / CF1043) sandwiched between crossed polarizers was added to the laser chain, after the final amplifier and before the compressor. This actively isolates the amplifiers from back-reflected seed pulses. The sub-10 nanosecond temporal gate window of the cell ensures that the forward going pulses pass through the system with minimal loss by 90-degree polarization rotation, while the back-reflected beams arriving after $>$60~ns  go through the cell in its `off' state. This results in rejection of their unchanged polarization state by the crossed polarizer at the entrance. To further protect the laser chain from any depolarized component of back-reflected pulse leaking through this system, all low extinction polarizers at the entrances and exits of the three manufacturer-implemented Pockels cells (two between the first and second amplifier, one between the pulse stretcher and first amplifier) were replaced with custom, high extinction, low group-velocity dispersion causing polarizers (Alpine Research Optics).

Several diagnostics were used to investigate various aspects of the laser-matter interactions and the production of secondary x-ray radiation. Radiation dose was measured using a Fluke Biomedical, Model 451P, Ion Chamber Survey Meter with a 25 keV low-energy detection threshold. A single-hit x-ray spectrometer with a custom collimator (Figs. \ref{fig:setup}b and \ref{fig:setup}c) isolated and measured the energies of backward-propagating x-rays and allowed an indirect assessment of primary backward-going electron energies. The apparatus was aligned along a line-of-sight to the target which `looks' through an aluminum vacuum flange, thinned to 1.7~mm at center, and the back of the OAP. Line-of-sight x-rays therefore have possible origins within the target itself, the OAP, or the aluminum flange. The x-ray spectrometer, an Amptek X-123 configured with a stack of three 3 mm x 3 mm CdTe detectors and a digitizer \cite{Amptek123_06}, saturates at 800~keV.  A custom collimator was designed and implemented to reduce the effective size of the detector and to shield the detector from non-line-of-sight photons (see diagram, Fig. \ref{fig:setup}b). This 102~mm thick collimator was machined from 936 bearing bronze (12\% Pb, 7\% Sn, and 81\% Cu).  Placement of the collimator and detector within a steel vacuum tube provided additional shielding and permitted the collimator (itself a potential source of non-line-of-sight photons) to be positioned 1.7~m in front of the detector. In order to estimate the measurement errors due to the detection of secondaries, the laser interaction chamber and single-hit spectrometer apparatus were modeled in MCNP \cite{MCNP}.  For 1.5~MeV mono-energetic, isotropic photon sources, MCNP predicts 1 scattered photon incident on the detector per 10 signal photons, a ratio which improves for lower energies. The placement of the collimator and the detector at 8.6 m and 10.3 m from the interaction region, respectively, results in a detection event rate less than one tenth of the laser repetition rate, and therefore a probability of double counts below $1\%$. For each x-ray spectrum, attenuation due to material in the line-of-sight was taken into account.

One of the experimental variables that has been found to strongly influence results from high intensity LPI experiments is the plasma scale length in front of the intended target caused by emission from the laser system preceding the main laser pulse, commonly referred to as `pre-pulse' \cite{mackinnon2001, ovchinnikov2013}. To determine the effect of pre-plasma in this experiment, the laser's temporal profile was measured on the nanosecond, picosecond, and femtosecond scales. A Thorlabs DET10A photodiode and 2.5 GHz Tektronix oscilloscope, a single shot autocorrelator \cite{salin1987}, and a scanning third-order cross correlator \cite{luan1993} were all used to characterize the Red Dragon's degree of pre-pulse. Similar to the nanosecond diagnostic outlined by \citet{lepape2009}, the photodiode was saturated to reveal low-signal nanosecond pre-pulse features. To increase dynamic range, multiple calibrated neutral density filters were successively removed, with measurements made at each level of saturation. The measurements were stitched together to create a single high-dynamic range temporal trace. The femtosecond-scale main pulse and its replicas were re-scaled according to the impulse response of the photodiode/oscilloscope system, a scaling confirmed by measurement of contrast of the amplified spontaneous emission (ASE) 600 ps prior to the main pulse with the third-order cross correlator.

The laser's nanosecond pre-pulse was controlled by manipulation of four Pockels cells at successive points in the laser amplification chain. The timing and physical angles of the Pockels cells were manipulated to achieve independent pre-pulse extinction ratios for each of the Pockels cells. For example, a good pre-pulse extinction after the 80 MHz laser oscillator will eliminate inherent femtosecond pre-pulse replicas arriving 12.5~ns, 25~ns,... prior to the main pulse. A deliberately poor pre-pulse extinction after the main amplifier will allow more ASE to arrive on target.

\begin{figure}
\centering
 \includegraphics[width=8.5cm]{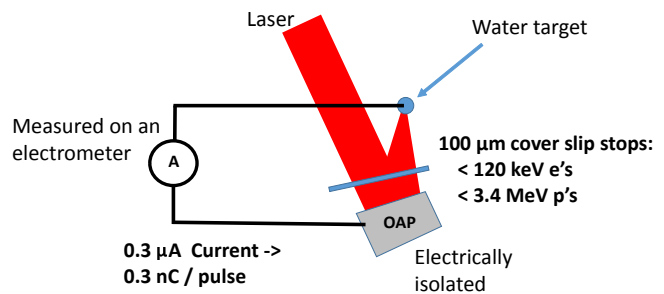}
\caption{\label{fig:charge}  Backward-directed charge is measured using the aluminum-substrate, gold-coated off-axis parabolic mirror (OAP) as a Faraday cup. Measurements are recorded as current on an electrometer, which is then converted to charge by dividing by the number of pulses per second (1000). A 100-\um{} thick glass cover slip is optionally inserted to block electrons with energies below 120~keV and protons with energies below 3.4~MeV. Measurements show $\sim$0.6~nC~/~pulse with no glass slide and $\sim$0.3~nC~/~pulse when the glass slide is inserted (with the laser in the `Uncleaned' condition of Fig.~\ref{fig:ppulse}). This indicates that each pulse produces nearly 0.3 nC of backward-directed electrons with energies above 120 keV.}
\end{figure}

\begin{figure*}
\centering
 \includegraphics[width=17cm]{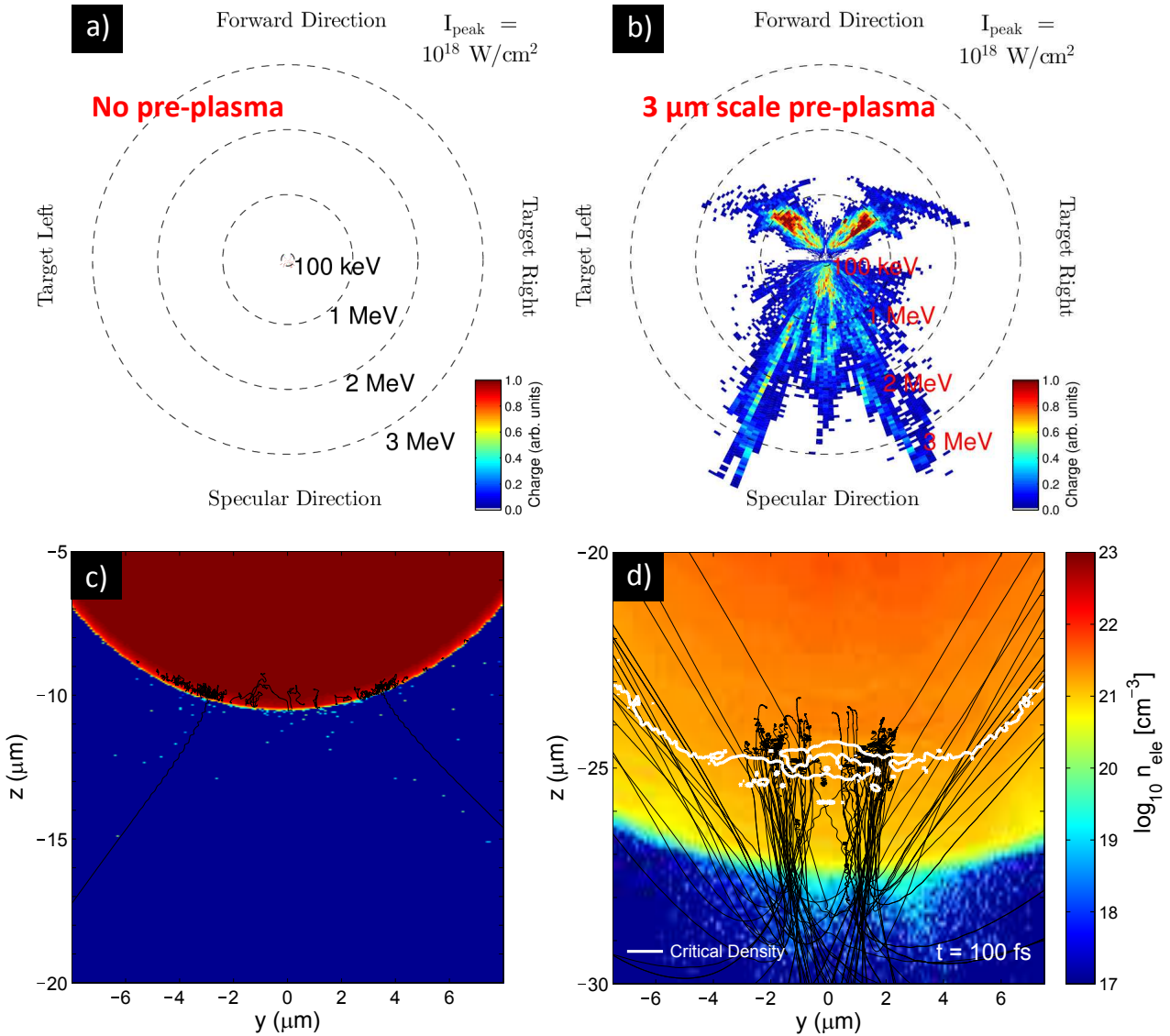}
 \vspace{-0.2cm}
 \caption{\label{fig:sim} Panels (a) and (c) show results from a 2D Particle-in-Cell (PIC) simulation with a sharp edge to the target, while panels (b) and (d) show PIC simulation results from a target with a 3~$\mu$m scale length. The upper panels show analysis of ejected electron macroparticles, where color indicates the total charge in each bin, distance from the origin indicates the electron energy, and the angle from the origin represents propagation direction. The lower panels show electron number density, with black lines highlighting trajectories of electron macroparticles. The simulations with a significant pre-plasma (right panels) show high energy (multiple-MeV) backward-propagating electrons whereas simulations without an appreciable pre-plasma (left panels) show very few escaping electrons.
}
\end{figure*}

To measure the number of backward-propagating electrons, the gold-coated, aluminum-substrate OAP was electrically isolated and used as a Faraday cup to collect electric charge (see Fig. \ref{fig:charge}). The collected charge was measured as an average current on a Keithley Instruments 610C solid-state electrometer, and this measurement was interpreted as charge per pulse after dividing by the number of laser-target interactions per second (1000). This method of obtaining charge per pulse was corroborated on a single-shot basis by analysis of the fast voltage trace observed using a 100 MHz oscilloscope in place of the electrometer. To determine the contribution of highly-energetic electrons to the measured charge, a 100~\um{} glass cover slip can be inserted into the laser beam path (see Fig. \ref{fig:charge}) to filter out the majority of backward-going $<$120~keV electrons and $<$3.4~MeV protons. The cover slip attenuates the laser by $\sim$15\% and slightly degrades the focal spot quality, reducing the laser intensity to \mbox{$\sim 7 \times 10^{17}$ W/cm$^2$}. 

\begin{figure*}
\centering
 \includegraphics[width=17cm]{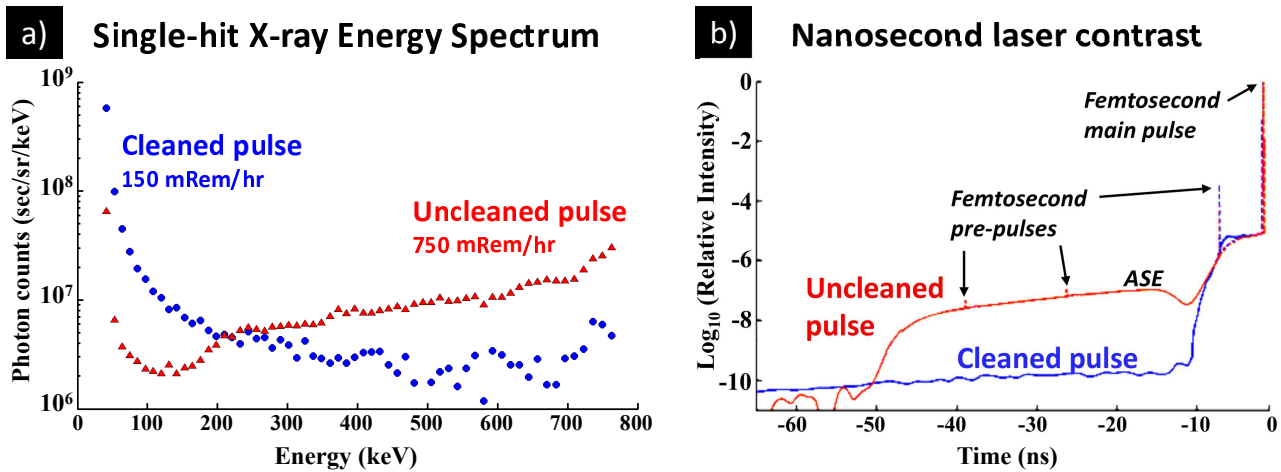}
  \vspace{-0.2cm}
 \caption{\label{fig:ppulse}a) X-ray spectra anti-parallel to laser incidence, produced from a 30~\um{} water column target for the two different nanosecond-timescale pre-pulse conditions highlighted in (b).  Panel (a) also reports the reading from a dosimeter placed immediately outside of the target chamber in the spectrometer's line-of-sight. A dose rate of 150~mRem/hr is recorded for the shorter Pockels cell gate timing, and 750~mRem/hr is measured with the increased pre-pulse. b) Photodiode measurements of the average ns-level contrast associated with the two spectra of panel (a). The solid lines show the laser contrast from two different Pockels cell timings and using seven calibrated filter conditions. The dashed lines rescale the photodiode response to the femtosecond main and pre-pulses.  Results are shown from the average of more than 1000 shots. The Pockels cell manipulated is located between the final amplifier and the compressor. For the `Uncleaned' case, we observe a smooth floor with spikes every 12.5~ns before the main pulse which is consistent with the period of the 80~MHz oscillator. The spike 6.2~ns before the arrival of the main pulse appears to instead arise from scattered light skipping a pass in our first amplifier. See Section~\ref{sec:Setup} for more details.}
\end{figure*}

\section{Simulations}\label{sec:Simulations}

The experiment's femtosecond-scale interactions were modeled with 2D(3v) Cartesian Particle-in-Cell (PIC) simulations using the LSP code \cite{welch2004} with the simulation setup as described in \citet{orban2015} except that the curvature of the water jet was included. The modeled laser peak intensity is \mbox{$1 \times 10^{18}$ W/cm$^2$} and the focal spot is Gaussian with 3.1~\um{} FWHM. As in the experiment, the laser polarization angle was perpendicular to the water column ($+y$ in Figs.~\ref{fig:sim}c and \ref{fig:sim}d) and the laser irradiated the target at normal incidence.  As in \citeauthor{orban2015} the spatial resolution was $\lambda / 32$ $\times$ $\lambda/32$ = $0.025 \, \mu$m $\times$ $0.025 \, \mu$m and time step was $\Delta t = 0.05 \,$fs. Both \citeauthor{orban2015} and the PIC simulations we present here use an implicit algorithm that avoids artificial heating and maintains good energy conservation in spite of the very high (close to solid) densities that exist in the simulation. Each cell with non-zero density was assigned 49 electron macroparticles, 49 singly ionized
oxygen macroparticles and 49 proton macroparticles. The initial temperatures were all set to 1 eV. All four simulation boundaries are `vacuum'; none are periodic or reflecting.

Fig. \ref{fig:sim}a presents an analysis of escaping electrons from a simulation with a sharp boundary at solid density which approximates the `Cleaned' pre-pulse of Fig. \ref{fig:ppulse}. Fig. \ref{fig:sim}b shows another simulation with an exponentially decreasing electron density having a scale length of 3.0~\um{}. In this simulation the conversion efficiency of laser energy into backward-going electrons $>$120 keV was found to be 3.0\%. For the sharp boundary the efficiency was essentially zero. In a qualitative way this corroborates experimental results shown in Fig.~\ref{fig:ppulse}. Note that PIC simulations with intermediate pre-plasma scale lengths (e.g. 1.5~\um{} which is not shown) have conversion efficiencies between 0 and 3.0\%. For example, simulations with a 1.5~\um{} scale length yield a 0.4\% conversion efficiency.

The lower panels of Fig.~\ref{fig:sim} show the electron number density of the target (both panels using the same colorbar) and some representative electron trajectories using black lines. Fig.~\ref{fig:sim}c shows the trajectories of electrons (of any energy) that move significantly in the sharp boundary simulations. This can be compared with Fig.~\ref{fig:sim}d which is derived from the simulation with the 3~\um{} pre-plasma. Fig.~\ref{fig:sim}d shows the trajectories of 100 electron macroparticles that leave the grid with $>$100~keV kinetic energy. Figs.~\ref{fig:sim}c and \ref{fig:sim}d illustrate visually what Figs. \ref{fig:sim}a and \ref{fig:sim}b indicate quantitatively, namely that far fewer electrons are accelerated in the sharp boundary simulation.

It is interesting to compare the electron trajectories in Fig. \ref{fig:sim}d with similar figures that appear in \citet{orban2015} (especially Fig.~5 of that study). In spite of the differences between the simulations performed in \citeauthor{orban2015} and the 3~\um{} pre-plasma simulation shown here (i.e. scale length, target curvature, spot size), there tends to be significant numbers of electrons coming from near or slightly above the critical density surface in both studies. This regime of electron acceleration was not the focus of \citeauthor{orban2015} but many of the same ingredients exist to accelerate these particles. Near the critical density, for example, the electrons still experience strong deflections from the evanescent standing wave fields. And much like the standing wave ejected electrons described in \citet{orban2015} and \citet{kemp2009} that originate at densities below critical, the electrons that are ejected from near or slightly above the critical density can be strongly accelerated away from the target by the reflected laser pulse. \citeauthor{orban2015} emphasized that laser absorption is at a minimum at intensities near \mbox{$10^{18}$ W/cm$^2$}. Indications from the PIC simulations shown here and in that study, and even from independent experimental work \cite{panasenko2010}, imply that the reflected laser pulse is of similar intensity as the incident pulse. Finally, charge separation effects in the pre-plasma caused by ponderomotive steepening \cite{estabrook1983} can help launch electrons aware from the target. These are all important reasons why large numbers of electrons are accelerated to super-ponderomotive energies.

\section{Results and Discussion}\label{sec:Results}

The backward-propagating x-ray spectrum reveals a peak energy exceeding the single-hit detector's saturation rejection limit of 800~keV (Fig. \ref{fig:ppulse}a, `Uncleaned' curve). The dominant x-ray source for the single-hit x-ray energy spectrum is likely the bremsstrahlung emission from high-energy backward-propagating electrons scattering off of atoms within the OAP. This interpretation is based upon several observations: the large number of electrons collected at the OAP, the demonstration of a Lanex-imaged electron beam (c.f. \citet{orban2015}), and the measurements of low isotropic x-ray dose emerging directly from the water jet itself (see Appendix).  Having established beyond reasonable doubt that the measured x-rays arise from electron bremsstrahlung, this implies source electrons with equal or higher energies than the measured x-ray energies. The 800 keV cutoff of the single-hit x-ray detector therefore implies the existence of $\sim$MeV energy electrons, which is well above the 110~keV ponderomotive energy scale (Wilks scaling \cite{wilks1992}) of the input laser and significantly higher energy than previous studies with similar experimental conditions \cite{uhlig2011}.

Our experimental results also show that when the nanosecond-scale pre-pulse is reduced through manipulation of laser Pockels cell gating, the backward-going x-ray spectrum (and by extension, the unmeasured backward-propagating electron spectrum) is modified substantially. The high energy components are suppressed and the radiation dose decreases from 750~mRem/hr to 150~mRem/hr despite a $<$5\% decrease in the total laser energy.

The measurements of electrons collected at the OAP provides a lower bound on the conversion efficiency from laser energy to energetic electrons.  Under conditions of significant pre-pulse and highly-energetic backward-going secondary x-ray radiation, the time-averaged electron current measured on the OAP Faraday cup was 0.6~\textmu{}A. When the 100~\um{} glass slide was added, the measured value was reduced to 0.3~\textmu{}A (Fig. \ref{fig:charge}). Given the kHz repetition rate of the laser, these values correspond to 0.6~nC~/~pulse and 0.3~nC~/~pulse, respectively. No current was measured on the OAP when the laser was blocked. An analysis of electron stopping distance in glass indicates it will block most electrons below 120~keV. For a conservative lower bound estimate of the conversion efficiency, if we assume each electron that passes is 120~keV we obtain a lower bound of 1.5\% conversion from laser energy into $>$120~keV electrons.

This result can be compared to 2D PIC simulations. PIC simulations of Section~\ref{sec:Simulations} show a \mbox{$10^{18}$ W/cm$^2$} laser interacting with a 3.0 \um{} pre-plasma scale length water target exhibiting 3.0\% conversion efficiency from laser energy into $>$120~keV backward-directed electrons. This simulation qualitatively confirms our experimental observations of significant backward-directed electrons with $\sim$MeV energies. PIC simulations with less extended pre-plasmas exhibited a lower conversion efficiency, qualitatively confirming experimental results with a `Cleaned' laser pre-pulse.

\citet{orban2015} used PIC simulations very similar to those employed here to explain the mechanism for accelerating these electrons in the water jet experiment. Relativistic electron energies are obtained through a standing-wave mechanism \cite{kemp2009} that injects electrons into the reflected laser pulse where they receive another energy boost. The standing wave pattern must exist in the pre-plasma in order for this injection to occur and electrons may be injected by the evanescent standing wave fields near the critical density surface. Eliminating the pre-pulse causes a steeper density gradient to exist, which reduces the number of electrons that experience the standing wave fields. The importance of normal incidence in this framework is that it serves to enhance the standing-wave electric and magnetic fields.

While it is remarkable that ultra-intense laser interactions at \mbox{$10^{18}$ W/cm$^2$} can produce such large amounts of backward-directed radiation, our results show that significant backward-going secondary radiation can occur when the laser is close to normal incidence and the reflectivity is high, as is the case in the experiments we describe in this paper.

\section{Conclusion}\label{sec:Conclusion}

In conclusion, we have demonstrated generation of $\sim$MeV electrons in the direction opposite laser propagation following the relativistic interaction at normal incidence of a 3~mJ, \mbox{$10^{18}$ W/cm$^2$} short pulse laser with a flowing 30~\um{} diameter water column target. We have measured MeV secondary x-rays created through bremsstrahlung collisions of hundreds of pC of electron charges per shot on the aluminum-substrate OAP. The measured doses in our experiment were substantial, and necessitated radiation protection for the personnel. The backward-going radiation spectrum was found to be sensitively dependent on the presence of a pre-plasma. These experimental results have been corroborated in 2D PIC simulations that show backward-going $\sim$MeV electron generation in similar number.

\pagebreak

\appendix*
\onecolumngrid

\section{Isotropic X-rays From Target}\label{sec:Isotropic}

\begin{figure}[H]
\centering
 \includegraphics[width=17cm]{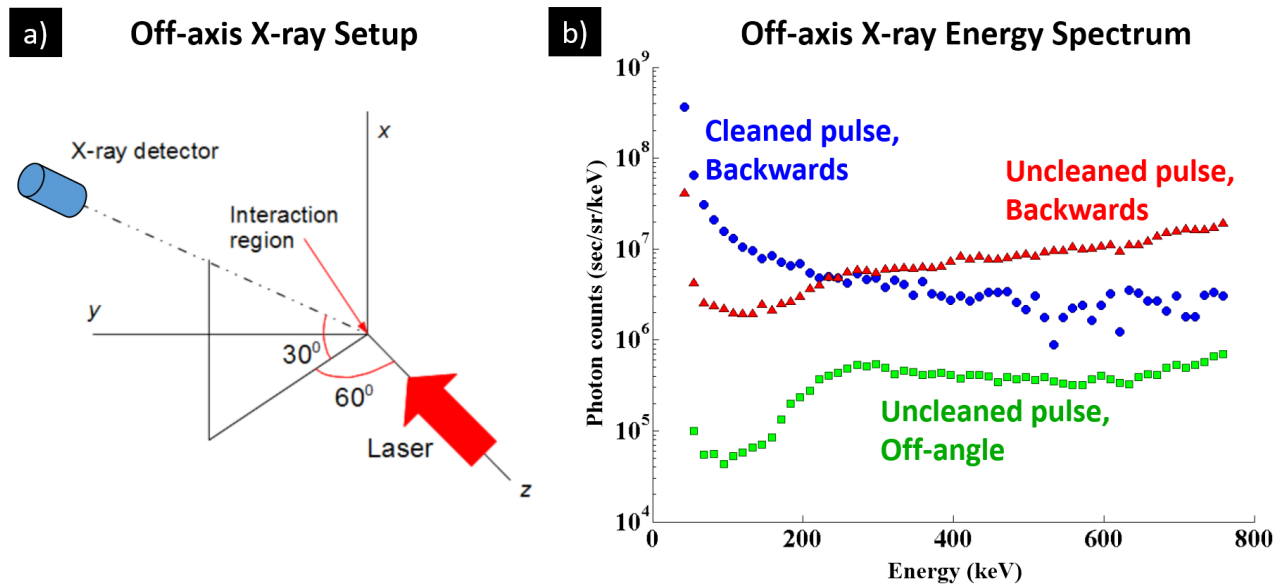}
 \vspace{-0.2cm}
 \caption{\label{fig:offangle} a) The `off-axis' setup is diagrammed, with the detector positioned 60 degrees to the side and 30 degrees upward, with respect to the laser normal axis. The collimator and single-hit x-ray detector are respectively placed 1.5~m and 2.7~m from the target to keep the double count rate below 1\%. b) The backward-going line-of-sight x-ray spectra for `Cleaned' and `Uncleaned' pre-pulse conditions, shown first in Fig. \ref{fig:ppulse}a, are re-shown for comparison with the `Uncleaned' off-axis line-of-sight x-rays (green spectrum). }
\end{figure}

In an alternate configuration to the `directly backward' x-ray spectrum measurements of Fig. \ref{fig:ppulse}a, the single-hit x-ray detector and collimator were set up `off-axis' to determine isotropic contributions originating from the water target itself. In this configuration (Fig. \ref{fig:offangle}a), only the target and a 1~mm thick aluminum vacuum flange were in the collimator line-of-sight. X-rays originating in the OAP were deliberately excluded from the line-of-sight and additionally obscured by a 51~mm thick lead block. For the same pre-pulse condition, the off-axis x-rays originating in the target and/or aluminum flange were at least an order of magnitude less numerous than the backward-going x-rays originating in the target, OAP, or aluminum flange (see Fig. \ref{fig:offangle}b). This suggests that the isotropic x-ray contribution from the water jet target to the backward-going x-ray spectrum is low.

\begin{acknowledgments}
This research was sponsored by the Quantum and Non-Equilibrium Processes Division of the Air Force Office of Scientific Research, under the management of Dr. Enrique Parra, Program Manager. Simulations were performed using the DOD HPC machines Spirit and Garnet and using storage at the Ohio Supercomputer Center.
\end{acknowledgments}


\bibliography{bibfile}


\end{document}